          [2001/04/25 1.1 (PWD)]
\documentclass[a4paper,twocolumn]{esapub} % European paper
\usepackage{natbib}
\usepackage{epsfig}

\title{The INTEGRAL Burst Alert System: Results and Future Perspectives}
\author{S. Mereghetti}
\author{D. G\"{o}tz}
\affil{ Istituto di Astrofisica Spaziale e Fisica Cosmica, Sezione
di Milano  G. Occhialini, CNR, Italy}
\author{J. Borkowski}
\affil{CAMK, Warsaw, Poland}
\author{M. Beck}
\affil{INTEGRAL Science Data Center, Versoix, Switzerland}
\author{A. von Kienlin}
\affil{MPE, Garching, Germany}
\author{N. Lund}
\affil{Danish Space Research Institute Copenhagen, Denmark}

\begin{document}

\keywords{Gamma-ray Bursts, transients}

\maketitle

\begin{abstract}
The INTEGRAL Burst Alert System (IBAS) is the software for real
time detection of Gamma Ray Bursts (GRBs) and the rapid
distribution of their coordinates. IBAS has been running almost
continuously at the INTEGRAL Science Data Center since the
beginning of the INTEGRAL mission, yielding up to now accurate
localizations for 12 GRBs detected in the IBIS field of view. IBAS
is able to provide error regions with radii as small as 3
arcminutes (90\% c.l.) within a few tens of seconds of the GRB
start. We present the current status of IBAS, review the results
obtained for the GRBs localized so far, and briefly discuss future
prospects for using the IBAS real time information on other
classes of variable sources.
\end{abstract}

\section{Introduction}

A new era in the study of Gamma-ray Bursts (GRBs) started with the
\textit{BeppoSAX} observations leading to the discovery of their
X--ray, optical and radio afterglows \citep{costa,vanpa,frail}.
The great progress which occurred in the last few years in our
understanding of GRBs has been possible thanks to extensive
multi-wavelength observations of these unpredictable and rapidly
fading events. In this respect, a quick derivation and
distribution of accurate sky positions for GRBs is crucial. Here
we review the contribution in this field obtained during the first
18 months of the INTEGRAL mission. We concentrate on the GRBs
observed within the field of view of the IBIS instrument
\citep{ibis}. Bursts observed with the SPI Anticoincidence Shield
(ACS) are described elsewhere in these proceedings \citep{acs}.

Thanks to its 72 hours orbit, the INTEGRAL satellite is in
continuous contact with the ground stations during the
observations. This has allowed us to implement a ground-based
software, the INTEGRAL Burst Alert System (IBAS), for the search
in near real time of GRBs \citep{ibas}. The IBAS software and its
current performances are briefly described in Section 2. In
Section 3 we summarize the main results on the twelve GRBs
observed to date in the field of view of the INTEGRAL instruments.
Finally, in Section 4 we describe the  IBAS capability  to provide
real time information also on other classes of transient sources.

\section{IBAS description and performances}

A detailed description of  IBAS  is given in \citet{ibas}. Here we
briefly remind the most salient features of the system.

As mentioned above, the search for GRBs is done on ground, at the
INTEGRAL Science Data Centre \citep{isdc}. In fact, no on-board
triggering system is present on INTEGRAL and the operating modes
of the instruments do not change when a GRB occurs. Since, under
nominal conditions, the telemetry data reach the ISDC without
important delays, the IBAS programs can run in near real time.
Such a ground based system offers some advantages with respect to
systems operating on board satellites, e.g. a larger computing
power and more flexibility for software and hardware upgrades. In
fact, in the course of the first year after the launch of INTEGRAL
several changes and additions have been done to the IBAS programs.
The current configuration is based on two different methods to
look for GRBs in the data from the IBIS lower energy detector
ISGRI \citep{isgri}.

In the first method the overall counting rate is monitored to look
for  significant excesses with respect to a running average of the
background, in a way similar to traditional triggering algorithms
used on-board previous satellites. Several different energy ranges
and integration times (from 2 ms to 5.12 s) are sampled in
parallel. A rapid imaging analysis is performed only when a
significant counting rate excess is detected. Imaging allows to
eliminate many false triggers caused, e.g. by instrumental effects
or background variations that do not produce a point source in the
reconstructed sky images. The second method is entirely based on
imaging. Images of the sky are continuously produced (integration
times of  10, 20, 40 and 100 s) and compared with the previous
ones to search for new sources.

The GRB positions derived by IBAS are delivered via Internet to
all the interested users. For the GRBs detected with high
significance, this is done immediately by the software which sends
\textit{Alert Packets} using the UDP transport protocol. In case
of events with lower statistical significance, the alerts are sent
only to the members of the IBAS Localization Team, who perform
further analysis and, if the GRB is confirmed, can distribute its
position with an \textit{Off-line Alert Packet}.

\begin{figure}[ht!]
\begin{center}   %  \vspace{2cm}
%      \hspace{0cm}\psfig{figure=fov.ps,width=7.5cm,angle=00  }
      \hspace{0cm}\psfig{figure=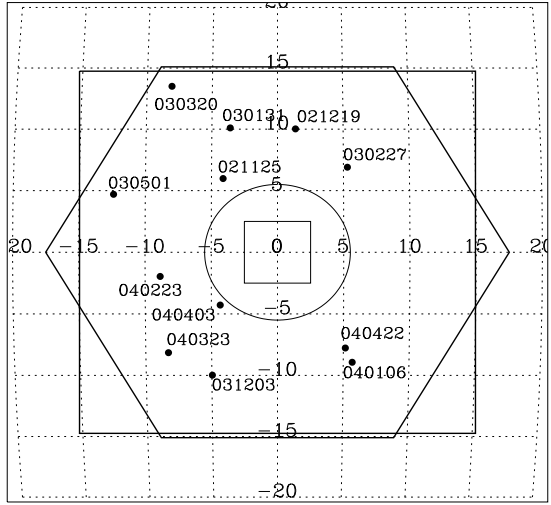,width=7.5cm,angle=90  }
      \caption[]{Positions of the GRBs detected in the field of view of the INTEGRAL instruments:
      IBIS (large square), SPI (hexagon), JEM-X (circle) and OMC (small square).
      The scale is in degrees.}
         \label{fov}
         \end{center}
   \end{figure}

%--------------------------------------------------------
\begin{table*}
\begin{center}
\caption[]{IBAS performance in   GRB localization}
\begin{tabular}{ccccc}
\hline \noalign {\smallskip}
GRB & Approximate &    Delay$^{a}$ in       & External delivery &  References \\
    &  duration        & position distribution   & of IBAS              &    \\
    & [s]      &   internal/public      & \textit{Alert Packets} &   \\
\hline \noalign {\smallskip}
021125  &  25  & --$^{b}$ / 0.9 days & OFF &  \citet{021125D} \\
021219  &   6  & 10 s / 5 hr         & OFF &  \citet{021219D} \\
030131  &  150 & 21 s / 2 hr         & ON  &  \citet{030131D} \\
030227  &  20  & 35 s / 48 min       & OFF &  \citet{030227D} \\
030320  &  50  & 12 s /  6 hr        & ON  &  \citet{030320D} \\
030501  &  40  & 30 s / 30 s         & ON  &  \citet{030501D} \\
031203  &  30  & 18 s / 18 s         & ON  &  \citet{031203D} \\
040106  &  60  & 12 s / 12 s         & ON  &  \citet{040106D} \\
040223  &  250 & 210 s / 210 s       & ON  &  \citet{040223D} \\
040323  &  20  & 30 s / 30 s         & ON  &  \citet{040323D} \\
040403  &  35  & 21 s / 21 s         & ON  & \citet{040403D} \\
040422  &   8  & 17 s / 17 s         & ON  & \citet{040422D} \\
 \noalign {\smallskip}
\hline \label{tab:spec}
\end{tabular}
\end{center}
$^{a}$ Computed from  the GRB start time.

$^{b}$ The IBAS \textit{Detector Programs} were in idle mode owing
to the limited telemetry allocation for IBIS/ISGRI during this
observation.

\end{table*}
%----------------------------------------------------------------

The first two months of operations after the INTEGRAL launch were
devoted to the optimization of the IBAS parameters. Some changes
in the algorithms were also required to adapt them to the
in-flight data characteristics. Delivery of the \textit{Alert
Packets} to the external clients started on January 17, 2003.
Since then it has always been enabled, except during the first
calibration campaign on the Crab Nebula (12-28 February 2003), and
a few very short interruption (few hours) due to maintenance
reasons.

Up to now (April 2004), twelve GRBs have been discovered in the
field of view of IBIS. Figure \ref{fov} shows their positions in
the fields of view of the INTEGRAL instruments. All of them were
at off-axis angles too large to be seen with the OMC and JEM-X
instruments.

\begin{figure*}[ht!]
\begin{center}
    %  \vspace{2cm}
%      \hspace{0cm}\psfig{figure=delays.ps,width=14cm,angle=90}
      \hspace{0cm}\psfig{figure=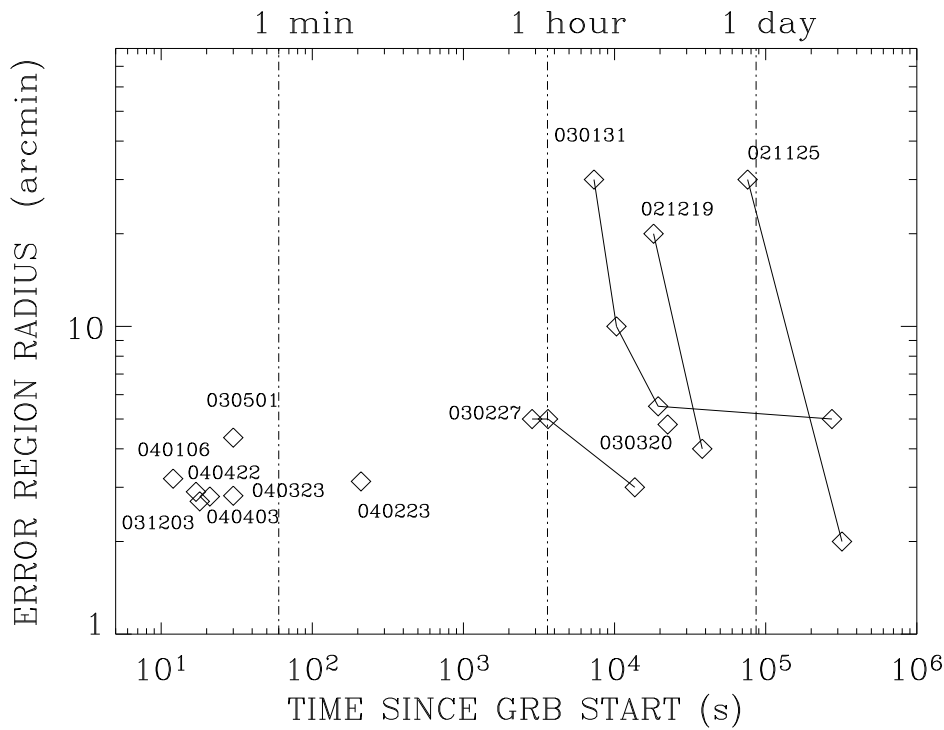,width=14cm,angle=90}
      \caption[]{Accuracy versus time delay for the 12 GRBs
      localized by IBAS. The figure  refers  to the external distribution of
      GRB coordinates with IBAS Alert Packets and GCN circulars.}
         \label{delays}
\end{center}
\end{figure*}

\begin{figure*}[ht!]
\begin{center}
    %  \vspace{2cm}
      \hspace{0cm}\psfig{figure=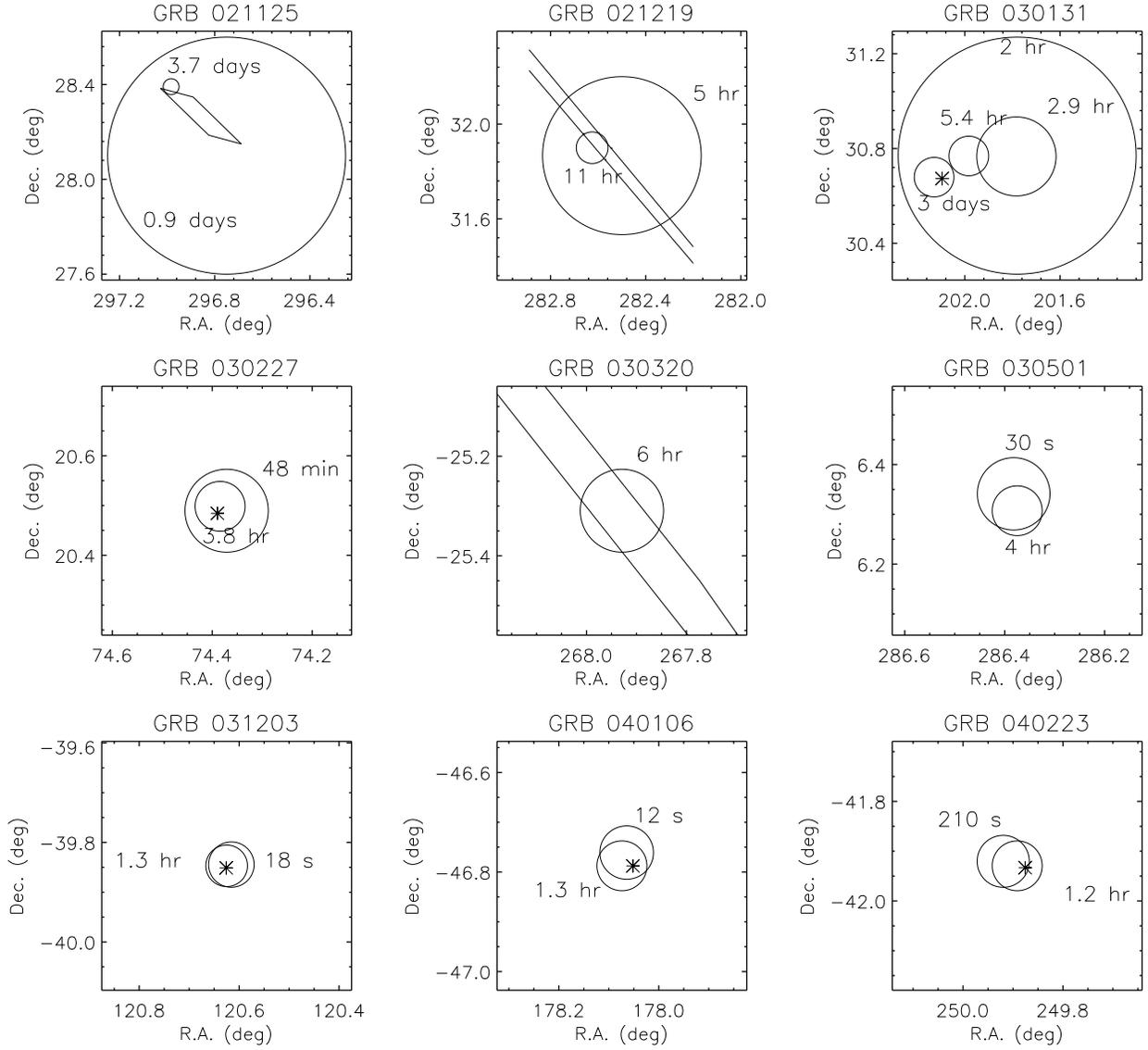,width=17cm,angle=90
      }
      \caption[]{Error regions   for the first nine  GRBs in the field of view
      of IBIS. The corresponding time delays are also indicated.
      Note the different scale of the three panels in the top line
      (1$^{\circ}\times$1$^{\circ}$ versus 0.5$^{\circ}\times$0.5$^{\circ}$).
      The parallelogram and the straight lines indicate error regions independently
      derived with the IPN. The asterisks mark the positions of the
      X-ray and/or optical afterglows. The localization of GRB 030131 was complicated
      by the fact that this burst was detected while  the satellite was performing a slew.}
         \label{erboxall}
\end{center}
\end{figure*}

The time and accuracy performances of the IBAS localizations for
these bursts are  summarized in Table 1 and illustrated in Figs.
\ref{delays} and \ref{erboxall}.  Note that at the beginning of
the mission the in-flight instrument misalignment was not
calibrated yet. Therefore,  error radii as large as 20$'$ or 30$'$
were given. The systematic uncertainties could be reduced in the
following months, leading to smaller error regions.

The time delay in the distribution of coordinates results from the
sum of several factors. First of all there is a delay on board the
satellite, which is variable and  depends on the instrument. For
IBIS/ISGRI data the average delay is  about 5 s. Signal
propagation to the ground station is negligible (maximum $\sim$0.6
s), but some time is required before the data are received at the
ISDC. This is on average 3 s when the ESA   ground station in Redu
(Belgium) is used, or 6 s when the NASA Goldstone ground station
is used. The time to detect the GRB depends on the algorithm which
triggers. The delay between the trigger time and the GRB onset is
of course dependent on the intensity and  time profile of the
event. The IBAS simultaneous sampling in different timescales
should ensure a small delay in most cases, however in practice a
minimum of $\sim$3 s is required to accumulate an image with
enough statistics. Finally, the conversion to sky coordinates,
comparison with list of known variable sources, \textit{Alert
Packet} construction and delivery require less than about 2 s. Of
course,  the above numbers assume nominal condition, i.e. no
telemetry gaps, no saturation of the allocated telemetry, no
missing auxiliary data files, etc...

As can be seen in Fig. \ref{delays} all the burst detected by IBAS
after April 2003 had very small error regions distributed within a
few tens of seconds, often while the gamma-ray emission was still
visible. Such a combination of high speed and small error region
was never achieved before. Note that the 210 s delay in the
localization of GRB 040223 was due do the particular light curve
shape of this burst lasting about 4 minutes and with the brightest
peak at the end.

%--------------------------------------------------------
\begin{table*}[ht!]
\begin{center}
\caption[]{Properties of GRBs detected with IBAS}
\begin{tabular}{cccccccccc}
\hline \noalign {\smallskip}
GRB      &  Peak Flux         & Peak Flux     & Fluence & Power law &Ref.$^{b}$ & Afterglow & Ref.$^{b}$ \\
         &  (20-200 keV)      & (20-200 keV)  & (20-200 keV)   & photon index$^{a}$ &  & & \\
       & [ph cm$^{-2}$s$^{-1}$] & [erg cm$^{-2}$s$^{-1}$] & [erg cm$^{-2}$] &        &  & & \\
\hline \noalign {\smallskip}
021125   & 22  &  2 $\times10^{-6}$  &7.4$\times 10^{-6}$ & 2.2/3.7 & (1)  & -- & \\
021219   & 3.7 & 3.5$\times10^{-7}$  &  9$\times 10^{-7}$ & 1.3$\rightarrow$2.5  & (2) & -- & \\
030131   & 1.9 & 1.7$\times10^{-7}$  &  7$\times 10^{-6}$ & $\sim2$& (3) & opt. & (3,4) \\
030227   & 1.1 & 1.6$\times10^{-7}$  &7.5$\times 10^{-7}$ & 1.9 & (5) & opt./X & (6,5) \\
030320   & 5.7 & 5.4$\times10^{-7}$  &1.1$\times 10^{-5}$ & 1.3$\rightarrow$1.9 & (7) & -- &  \\
030501   & 2.7 &   3$\times10^{-7}$  &  3$\times 10^{-6}$ & 1.75 & (8)  & -- & \\
031203   & 1.2 & 1.3$\times10^{-7}$  &  $^{c}$    &    $^{c}$        & (9) & radio/opt./X &  (10,11,12) \\
040106   & 0.6 & 6.5$\times10^{-8}$  &  $^{c}$    &    $^{c}$        & (13) & opt.?/X &  (14,15) \\
040223   & 0.4 &   3$\times10^{-8}$  &  $^{c}$    &    $^{c}$        & (16) & X &   (17) \\
040323   & 1.7 & 2.2$\times10^{-7}$  &  $^{c}$    &    $^{c}$        & (18) & opt.? & (19)  \\
040403   & 0.4 &   3$\times10^{-8}$  &  $^{c}$    &    $^{c}$        & (20) & &   \\
040422   & 2.7 & 2.5$\times10^{-7}$  &  $^{c}$    &    $^{c}$        & (21) & &   \\
 \noalign {\smallskip} \hline \label{tab:spec}
\end{tabular}
\end{center}
$^{a}$ The two values for GRB 021125 are for the ranges 20-200 keV
(ISGRI) and 170-500 keV (PICsIt). The arrow indicates time
evolution.

$^{b}$ References: (1) \cite{021125P}; (2) \citet{021219P}; (3)
\citet{030131P}; (4) \citet{030131O}; (5) \citet{030227P}; (6)
\citet{030227O}; (7) \citet{030320P}; (8) \citet{030501P}; (9)
\citet{031203D}; (10) \citet{031203R}; (11) \citet{031203O}; (12)
\citet{031203X}; (13) \citet{040106D}; (14) \citet{040106O}; (15)
\citet{040106X}; (16) \citet{040223D}; (17) \citet{040223X}; (18)
\citet{040323D}; (19) \citet{040323O}; (20) \citet{040403D}; (21)
\citet{040422D};

$^{c}$ Results not yet published

\end{table*}
%----------------------------------------------------------------

\begin{figure}[ht!]
\begin{center}
    %  \vspace{2cm}
      \hspace{0cm}\psfig{figure=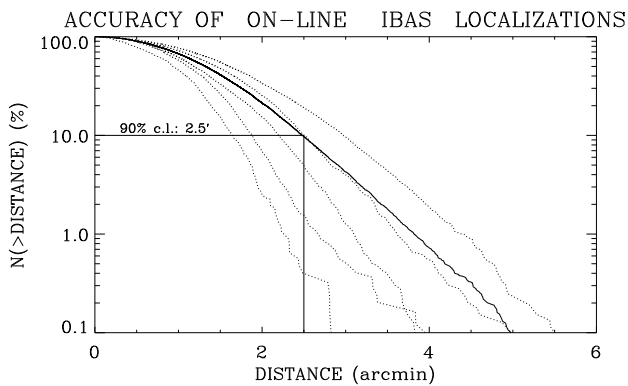,width=8.5cm,angle=00
      }
      \caption[]{Integral distribution of the angular difference between the true
      coordinates of known sources and those derived on-line by IBAS. The solid
      line is the total distribution  all the triggers.
      The dotted lines refer to triggers with  signal to noise
      values in different ranges (from top to bottom: 8-10, 10-12,
      12-15, 15-20, $>$20). As expected the localization accuracy
      increases for higher signal to noise triggers.}
        \label{acc}
  \end{center}
   \end{figure}

\begin{figure}[ht!]
  \begin{center}  %  \vspace{2cm}
      \hspace{0cm}\psfig{figure=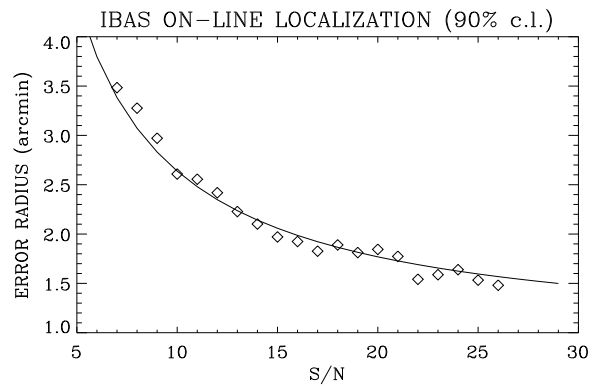,width=8.5cm,angle=00
      }
      \caption[]{Error on the coordinates derived by the on-line IBAS imaging programs
       as a function of the signal to noise ratio. The diamonds are experimental
       data derived from the observation of known sources. The line is an
       eye-fit with a simple analytical law.}
         \label{acc90}
         \end{center}
   \end{figure}

The accuracy of the localizations derived by the IBAS on-line
programs (and distributed in the automatic \textit{Alert Packets})
can be estimated based on the IBAS triggers caused by known
sources. Figure \ref{acc} shows the distribution of the
differences between the true and derived coordinates based on
$\sim$24,000 triggers from Sco X-1, Cyg X-1, Vela X-1 and other
sources with well known positions. From the curves shown in Fig.
\ref{acc} we can estimate the location accuracy as a function of
the source signal to noise ratio (S/N). This is reported in Fig.
\ref{acc90}, where it can be seen that the 90\% c.l. error radius
is smaller than 2.5$'$ for S/N$>$10 (the current threshold for the
automatic delivery of \textit{Alert Packets} containing the GRB
coordinates is S/N=8). Note that the above discussion refers to
the on-line imaging analysis which is based on simplified
algorithms. In general, the GRB error regions can be further
reduced by the more sophisticated interactive analysis performed
off-line.

\begin{figure*}[ht!]
\begin{center}
 %     \vspace{2cm}
      \hspace{0cm}\psfig{figure=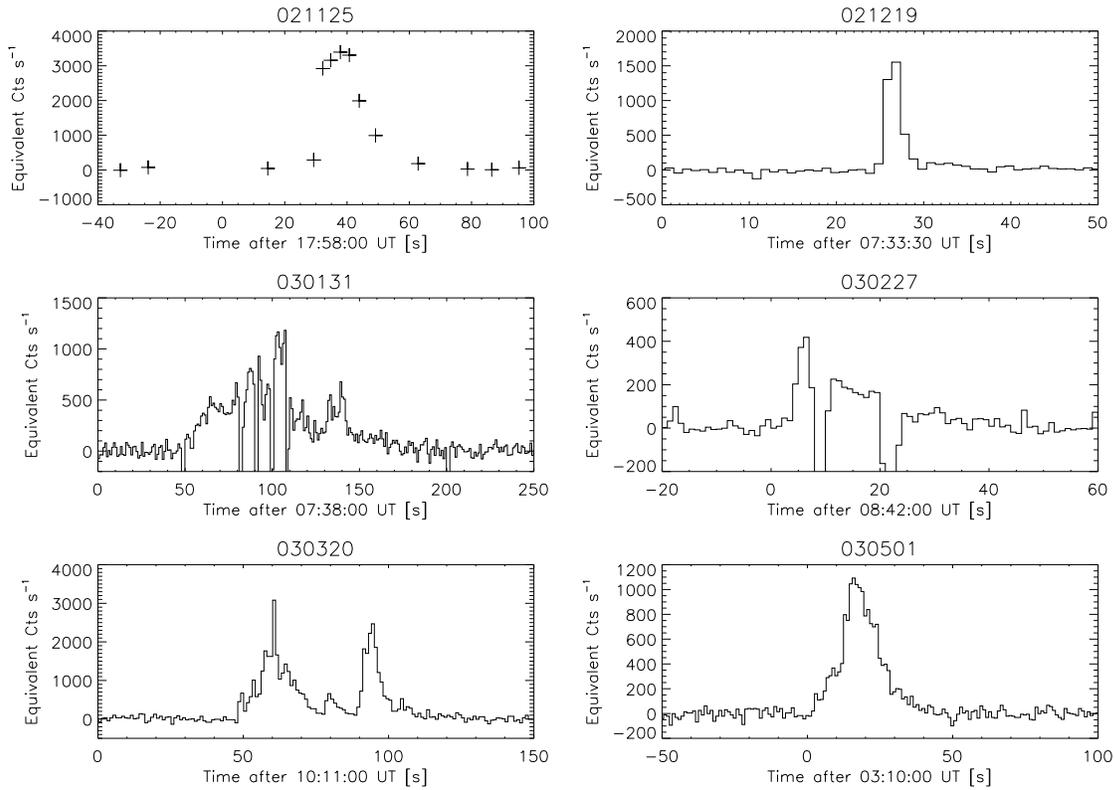,width=15cm,angle=00}
      \caption[]{IBIS/ISGRI light curves in the 20-300 keV energy range of
       the first six GRBs detected by IBAS. The background has been subtracted, and,
       except for GRB 030131 which was detected while the satellite was slewing, the
       light curves have been extracted from pixels illuminated by more than
       50\%  of their area. The count rates have been converted to their equivalent on-axis
       values. The gaps in the light curves of GRB 030131 and GRB 030227 are due
       to satellite telemetry saturation. All the light curves are binned at 1 s,
       except for GRB 021125, where most of the time bins are shorter due to
       the telemetry gaps (see \citet{021125P} for details).        }
         \label{bigplot}
\end{center}
\end{figure*}

\begin{figure}[ht!]
\begin{center} %     \vspace{2cm}
%      \hspace{0cm}\psfig{figure=pos.ps,width=7.5cm,height=5cm,angle=00}
      \hspace{0cm}\psfig{figure=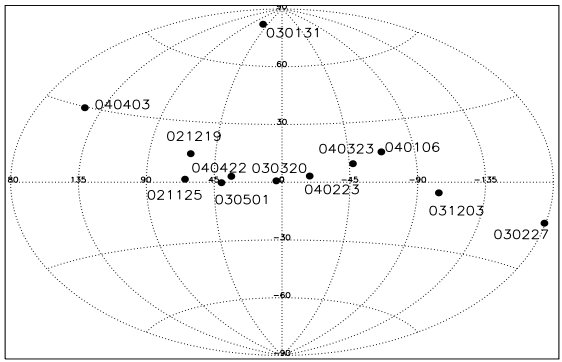,width=7.5cm,height=5cm,angle=90}
      \caption[]{Positions in Galactic coordinates of the twelve GRBs
      localized  so far by INTEGRAL. The large fraction of bursts  at low
      Galactic latitude reflects the non-uniform sky exposure obtained by
      INTEGRAL.}
         \label{pos}
         \end{center}
   \end{figure}

\begin{figure}[ht!]
\begin{center}
    %  \vspace{2cm}
%      \hspace{0cm}\psfig{figure=logNlogP.ps,width=8.5cm,angle=0}
      \hspace{0cm}\psfig{figure=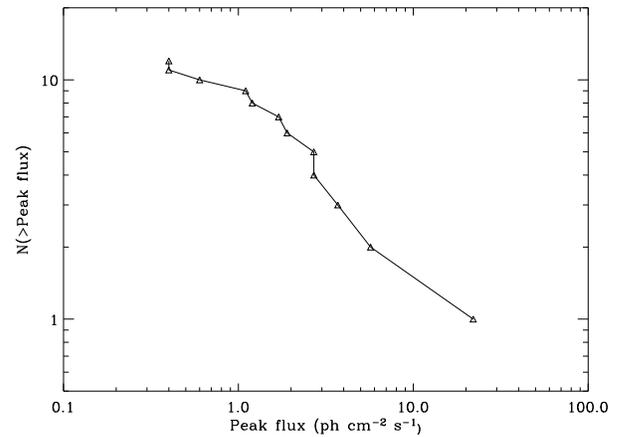,width=8.5cm,angle=90}
      \caption{Log N - Log P curve for the 12 bursts seen with IBAS.}
       \label{lognlogp}
\end{center}
   \end{figure}

\begin{figure*}[ht!]
\begin{center}
    %  \vspace{2cm}
%      \hspace{0cm}\psfig{figure=T90_Pflux4.ps,width=14cm,angle=90     }
      \hspace{0cm}\psfig{figure=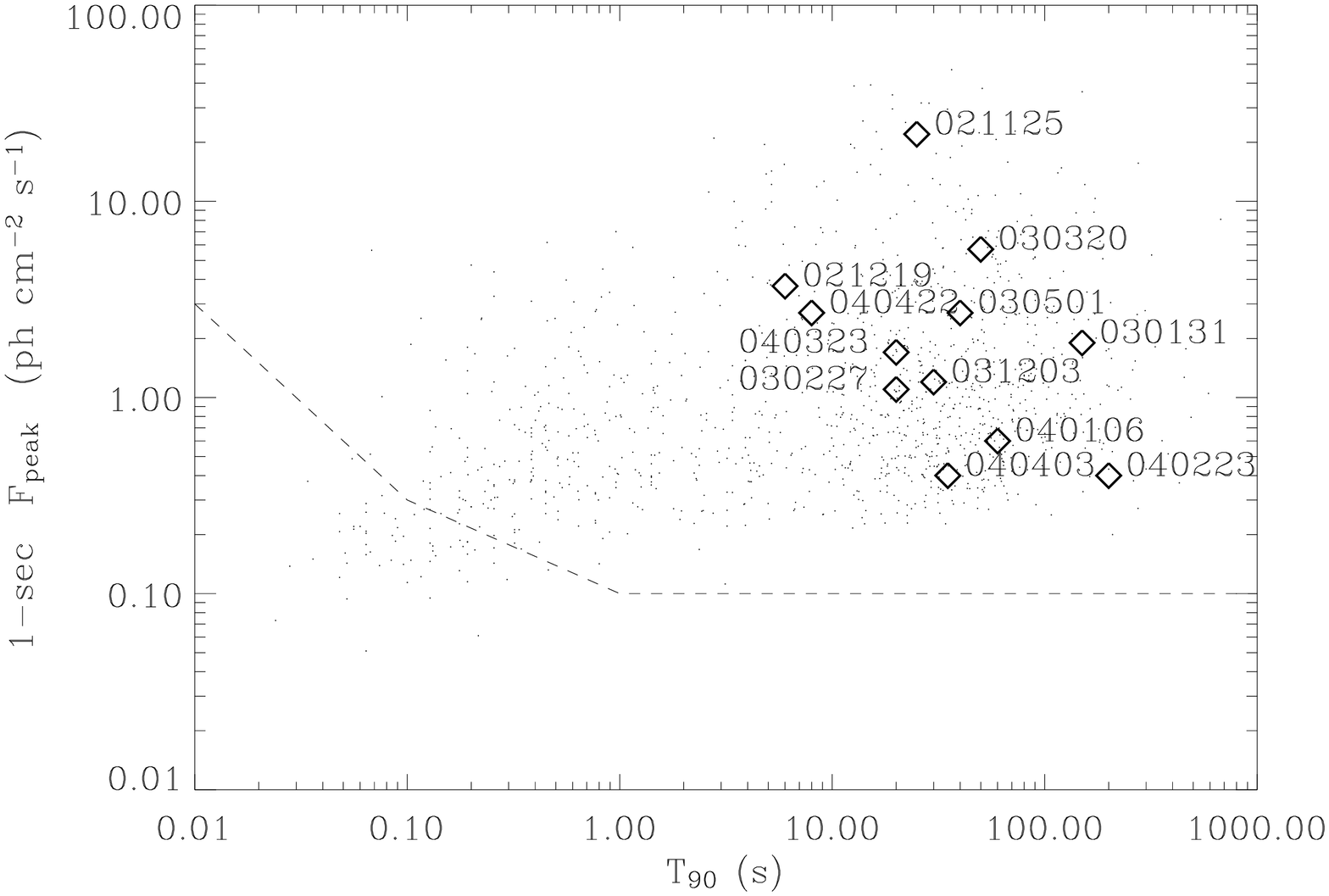,width=14cm,angle=90     }
      \caption[]{Peak flux versus duration for the 12 GRBs detected by INTEGRAL
      compared to the corresponding values of the BATSE catalog.
      The dashed line is an estimate of the on-axis sensitivity
      (of course, the exact value depends on the spectrum and
      time profile of the GRB).}
       \label{t90}
       \end{center}
   \end{figure*}

\section{Results on Gamma-ray bursts}

The main properties of the twelve INTEGRAL GRBs  are summarized in
Table 2. In Fig. \ref{bigplot} we show the light curves of the
first six bursts, as measured with the IBIS/ISGRI instrument. The
results for these six bursts, including their spectral analysis,
have been published in the references given in Table 2. Most of
the spectral information for these bursts is  based on IBIS/ISGRI
data. Although the nominal range of the instrument is 15 keV - 1
MeV, the coverage above $\sim$200-300 keV is actually limited by
the small statistics. Thus most of the spectra are well described
by a power-law over the $\sim$15-200 keV range (see photon index
values in Table 2). Only for GRB 030131 a curved spectrum gives a
better fit, with parameters of the Band model $\alpha$=1.4,
$\beta$=3 and E$_o$=70 keV \citep{030131P}. We remark that the
spectral response of IBIS in the partially coded field of view is
not completely calibrated yet. Therefore, the published GRB
spectra should be considered as preliminary.  No evidence for
spectral lines has been seen so far in the SPI spectra, which are
generally in good agreement with the IBIS/ISGRI ones.

Thanks to the good sensitivity of IBIS/ISGRI, it has been possible
to study the time evolution of the spectra even for relatively
faint bursts. A typical hard to soft variation has been clearly
seen in GRB 021219 and GRB 030320 \citep{021219P,030320P}, and
with lower significance in GRB 030227 and GRB 030131
\citep{030227P,030131P}.

The distribution of peak fluxes for the INTEGRAL bursts is shown
in Fig. \ref{lognlogp}, but of course the  small number of events
does not allow for the moment a meaningful interpretation of such
a LogN-LogP function.

As can be seen in Fig. \ref{t90} all the bursts detected so far
belong to the long duration class ($>$2 s). The optimization of
the IBAS programs and parameters for short bursts required some
time, leading, in the first months after the launch, to a bias in
favor of long GRBs. However, since the Summer of 2003 IBAS has
also a good sensitivity also for short events, as it has been
demonstrated by the real time detection of many weak bursts
lasting $\sim$0.1-0.2 s from SGR 1806-20 \citep{sgr1806}.
Therefore, we expect to obtain a rapid localization also for a
short burst in the coming months.

INTEGRAL spends most of the time pointing at Galactic targets. As
a consequence the majority of the detected bursts are at low
Galactic latitude (see Fig. \ref{pos}). This is not an ideal
situation for what concerns their follow-up observations at other
wavelengths. Nevertheless, thanks to the rapid dissemination of
their coordinates, X-ray and/or optical/IR afterglows have been
reported for six INTEGRAL GRBs.
%One of the most remarkable cases is that of GRB 031203 .....

\section{IBAS and other variable sources}

\begin{figure}[ht!]
\begin{center}
    %  \vspace{2cm}
      \hspace{0cm}\psfig{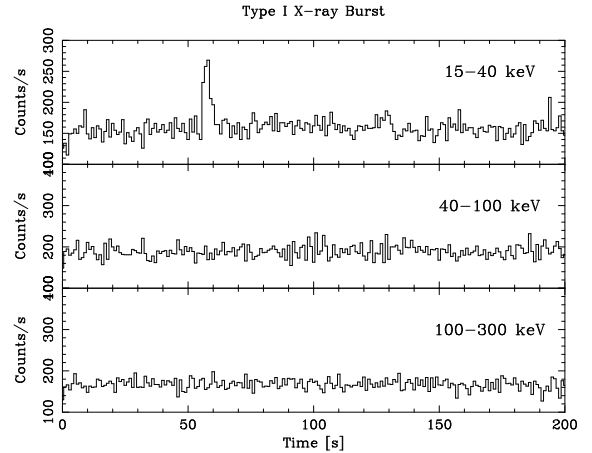}
      \caption[]{IBIS/ISGRI light curves  of a Type I burst from a Galactic bulge binary
      detected in real time by IBAS. The burst is clearly visible only
      in the lowest energy range.}
       \label{burst}
       \end{center}
   \end{figure}

The IBAS programs are sensitive to several kinds of
variable/transient events, in addition to GRBs. Most of the
triggers caused by bright and highly variable sources, like e.g.
Sco X-1 or Cyg X-1, represent just a disturbance to the main IBAS
task. However, some triggers are due to potentially interesting
phenomena.  These include type I and type II X-ray bursts from Low
Mass X-ray Binaries (see Fig. \ref{burst}), bursts from Soft
Gamma-ray Repeaters, outbursts from known and unknown transients.
As for GRBs, the information on the occurrence of these events is
derived in real time (i.e. within few tens of seconds). To avoid
alerting the community of users mainly interested in GRBs and to
comply with the INTEGRAL data rights rules, all the IBAS
\textit{Alert Packets} with derived coordinates consistent with
the positions in a list of known sources are not distributed
(except for Soft Gamma-ray Repeaters, whose alerts are being
distributed since January 2004).

We give here just two examples of interesting cases. One is the
new transient hard X-ray source IGR J17544--2619, discovered by
INTEGRAL in the Galactic bulge region. The second example is the
very high state reached by the high mass X-ray binary pulsar Vela
X-1 on November 28, 2003 \citep{velax1}. In both cases the IBAS
programs triggered in real time, while the public announcements
were given respectively 14 hours and 4 days later \citep{igr,
velax1D}. Rapid follow-up at other wavelengths would have provided
very useful information: the first outburst from IGR J17544--2619
(the one during which it was discovered) lasted only two hours,
while the 15-40 keV flux of Vela X-1 increased from 0.5 to 7 Crabs
during the first 5400 s before decreasing in the following hours.

In general, these ``secondary'' IBAS results are used to
complement the quick-look analysis performed at the ISDC. However,
the real time information which is potentially available through
IBAS is not fully exploited yet. We plan to implement these
capabilities in the future, possibly also including the use of
JEM-X data. Robotic optical/IR telescopes operating on ground, as
well as satellites with a rapid reaction time, such as Swift
\citep{swift}, could benefit from the IBAS alert messages to
perform follow-up observations of X-ray transients with an
unprecedented rapidity.

\section*{Acknowledgments}

This work has been partly funded by ASI. JB was supported by grant
2P03C00619p02 from KBN.

\end{document}